\documentclass[12pt]{article}
\usepackage{euscript,amsmath, amssymb, amsfonts}

\def\url#1{\expandafter\string\csname #1\endcsname}

\usepackage{url}

\newcommand{\be}{\begin{equation}}
\newcommand{\ee}{\end{equation}}

\begin{document}

\title{{SO}(4)-symmetry of mechanical systems with 3 degrees of freedom}

\author{Sofiane Bouarroudj\thanks{New York University Abu Dhabi,
Division of Science and Mathematics, P.O. Box 129188, United Arab
Emirates; \sf{sofiane.bouarroudj@nyu.edu}.}
 \ and S.E. Konstein
\thanks
{I.E.Tamm department of Theoretical Physics,
P.N. Lebedev Physical Institute of the Russian Academy of Sciences,
Leninskij prosp. 53, RU-119991 Moscow, Russia; \sf{konstein@lpi.ru}.}
}

\date{
}
\maketitle

\begin{abstract}
We answered the old question: does there exist a mechanical system with 3 degrees of freedom,
except for the Coulomb system, which has 6 first integrals generating the Lie algebra $\mathfrak{o}(4)$
by means of the Poisson brackets?
We presented a system which is not centrally symmetric,
but has such 6 first integrals.
We showed also that not every mechanical system with 3 degrees of freedom possesses such Lie algebra
$\mathfrak{o}(4)$.
\end{abstract}




\section{Introduction}

It is well-known (see, e.g., \cite{LLm}) that in the Coulomb field,
i.e., in the mechanical system with 3 degrees of freedom
($3d$ mechanical system)
with the Hamiltonian
\begin{equation}\label{H}
H=\frac{\mathbf{p}^2}{2}-\frac 1 r, \quad \mbox{ where } \quad \mathbf{p}^2:=\sum_{i=1,2,3} p_i^2,
\ r:=\left ( \sum_{i=1,2,3} q_i^2 \right )^{1/2},
\end{equation}
the symmetry group of canonical
transformations\footnote{Recall, that in Classical Hamiltonian Mechanics, the transformations of the phase space that preserve the Hamiltonian form of the Hamilton equations, whatever the Hamiltonian function is, are said to be \textit{canonical}.} has a subgroup isomorphic to $\mathrm{SO}(4)$
acting in the domain $H<0$. This
fact, found by V.~Fock \cite{Fock}, helps to explain the structure of
the spectrum of the hydrogen atom. Sometimes this symmetry is called \emph{hidden}.

An important property of this $\mathrm{SO}(4)$ is that the Casimirs of its Lie algebra $\mathfrak{o}(4)$ restore the Hamiltonian.
The Hamiltonian in Eq. (\ref{H}) describes, for example, the motion of two particles interacting via gravity,
and the motion of two charged particles with the charges of opposite sign.
The number of works investigating this
Hamiltonian is huge.\footnote{See, for example, \cite{Bulk, OhSi} and references therein.}
It is therefore astonishing that the literature does not give (at least, we could not find it)
the definite answer to a natural question:
``does there exist a mechanical system with
3 degrees of freedom, except for the Coulomb system, which has 6 first integrals generating the
Lie algebra $\mathfrak{o}(4)$?"
posed, e.g., in \cite{Mu, SW}.
Two different answers to the question were given fifty years ago:

1) Mukunda \cite{Mu} claimed that every mechanical system with $n$ degrees of freedom has
a subgroup of canonical transformations locally isomorphic to
$\textrm{O}(n+1)$.

2) Szymacha and Werle \cite{SW} claimed that there are no other mechanical system with the same property,
assuming that $\mathfrak{o}(4)$ contains the Lie algebra of spatial rotations of $\mathbb R^3$.

To prove that not any system with 3 degrees of freedom has $\mathrm{SO}(4)$ symmetry or at least
$\frak o (4)$ Lie algebra of the first integrals,
we offer a simple necessary condition for existence of $\mathfrak{o}(4)$ symmetry, see Section \ref{rank},
and in Section  \ref{omega} we give an example for which this condition is violated.

In Section \ref{sec6} we consider the Hamiltonian of a charged particle in an homogeneous electric field.
For this Hamiltonian, there exists a family of sextuples of first integrals such
that every sextuple generates (by means of the Poisson bracket) the Lie algebra $\mathfrak{o}(4)$.

To avoid misunderstanding, we should note that we consider the \emph{symmetry algebra}
(consisting of some set of the
first integrals)
of the systems,
not the Lie algebra of \emph{dynamical symmetry group}  introduced in \cite{Bar},
which is also called  \emph{noninvariance group},
see \cite{MRS}.

\section{Generalities (following \cite{Arnold})}\label{gen}

Recall the definition of the symmetry  group of canonical
transformations and its Lie algebra.

Let $H(q_i,\,p_i)$, where $i=1,2,3$,
be a Hamiltonian of some mechanical system.
We will also denote the whole set of the $q_i$ and $p_i$ for $i=1,2,3$
by $z_\alpha$, where $\alpha$ =1, \dots , 6.

Let the first integral $F$ of this system  be a real
function on some domain $U_F\subset \mathbb{R}^6$. Let $(q,p)\in U_F$; the case  $F=H$ is not excluded. Then
$F$ generates a
1-dimensional Lie group $\mathcal L_F$ of canonical transformations $(q,p)\mapsto (q^F(\tau|q,p),p^F(\tau|q,p))$
leaving the Hamiltonian $H$ and the domain $U_F$ invariant
if
\[
(q^F(\tau|q,p),p^F(\tau|q,p)) \in U_F\mbox{~~ for any } \tau\in \mathbb{R}.
\]
The transformations are defined by the relations
\begin{eqnarray}\label{hamfb}
&& \frac {dq_i^F}{d\tau}=\{q_i^F,\, F\}  = \frac {\partial F(q^F,\, p^F)}%
{\partial p_i^F} ,  \\
&& \frac {dp_i^F}{d\tau}=\{p_i^F,\, F\}
 = -\frac {\partial F(q^F,\, p^F)}%
{\partial q_i^F},\\
&& q_i^F(0|q,p)=q_i,\ \ p_i^F(0|q,p)=p_i,
\label{hamfe}
\end{eqnarray}
where $\{\cdot,\cdot\}$ is the Poisson bracket\footnote{The definition (\ref{PB}) has the opposite sign as compared with the one
given in \cite{LLm}, but coincides with the definition of the Poisson bracket
given in \cite{Appell, Ter-Haar, Gantmaher, Arnold}.}
 in $\mathbb{R}^6$:
\begin{equation}\label{PB}
\{F, G\}:=
\sum_{i=1,2,3}\left(\frac {\partial F} {\partial q_i} \frac {\partial G} {\partial p_i}
-\frac {\partial F} {\partial p_i} \frac {\partial G} {\partial q_i}\right)=
\sum_{\alpha,\beta=1, \dots, 6} \frac {\partial F} {\partial z_\alpha}
\omega_{\alpha\beta}
\frac {\partial G} {\partial z_\beta}.
\end{equation}
Here the symplectic form $\omega$ is of shape
$\omega=\left(
\begin{array}{cc}
0_3 & 1_3 \\
-1_3 & 0_3%
\end{array}%
\right) $, where $1_3$ and $0_3$ are $3\times 3$ matrices.

We call the transformations Eq. (\ref{hamfb}) -- (\ref{hamfe}) the \emph {Hamiltonian flow},
 generated by the
Hamiltonian $F$, and denote it  $\mathcal L_{F}$.

If a certain finite set of first integrals $\mathcal F =\{F_\alpha\mid \alpha=1,2,...,\}$
has the same domain $U$
invariant under the action of all Hamiltonian flows $\mathcal L_{F_\alpha}$,
then these flows generate a Lie group, the space of its  Lie algebra
being
generated by the set $\mathcal F$ by means of the bracket
(\ref{PB}).

\section{The case of the Coulomb field (following \cite{LLm})}\label{clmb}

Here we briefly consider  the mechanical system \eqref{H} \iftrue with the Hamiltonian
\begin{equation}\label{H1}
H=\frac{\mathbf{p}^2}{2}-\frac 1 r,\quad  \mbox{ where } \quad  \mathbf{p}^2:=\sum_{i=1,2,3} p_i^2,
\ r:=\left ( \sum_{i=1,2,3} q_i \right )^{1/2}\,.
\end{equation}\fi
This Hamiltonian has two well-known triples of first integrals: one consists of the
coordinates $L_i$ of the angular momentum vector,
the other one consists of the coordinates of the Runge-Lenz vector $R_i$,
defined in the domain
\[
U=\{z \in \mathbb R^6 \mid H(z)<0 \},
\]
or in any of the domains   $E_{\mathrm{min}}<H<E_{\mathrm{max}}<0$, by the formulas
\begin{eqnarray}
L_i &:=& \sum_{j,k=1,2,3} \varepsilon_{ijk} q_j p_k,    \\
R_i &:=& (-2H)^{-1/2} \left( \sum_{j,k=1,2,3} \varepsilon_{ijk} L_j p_k  +\frac{q_i}{r}\right),
\end{eqnarray}
where $\varepsilon_{ijk}$ is an anti-symmetric tensor such that $\varepsilon_{123}=1$.

These  first integrals satisfy the following commutation relations:
\begin{equation}\label{HF}
\begin{array}{ll}
\{H, L_i\}=0,\\
\{H, R_i\}=0,
\end{array}
\end{equation}
and
\begin{equation}\label{1}
\begin{array}{ll}
\{L_i, L_j\}=\sum_{k=1,2,3}\varepsilon_{ijk}L_k,\\
\{R_i, R_j\}=\sum_{k=1,2,3}\varepsilon_{ijk}L_k,\\
\{L_i, R_j\}=\sum_{k=1,2,3}\varepsilon_{ijk}R_k.
\end{array}
\end{equation}

Due to relations  (\ref{HF}) and by definition of the domain $U$, the later is invariant under
the action of Hamiltonian flows generated by the first integrals $L_i$ and $R_i$.

The relations  (\ref{1}) show that these first integrals generate the Lie algebra $\mathfrak{o}(4)$.

Since $\mathfrak{o}(4)\simeq \mathfrak{o}(3) \oplus \mathfrak{o}(3)$, we can introduce
two commuting triples of first integrals
\begin{equation}\label{o3}
\begin{array}{lcl}
G_i &:=& \frac 1 2 (L_i+R_i), \ \mbox{~where~~} i=1,2,3,\\
G_{3+i}&:=& \frac 1 2 (L_i-R_i), \ \mbox{~where~~}  i=1,2,3,
\end{array}
\end{equation}
satisfying the commutation relations
\begin{equation}\label{o3o3}
\begin{array}{lcl}
\{G_i, \, G_j\} &=& \sum_{k=1,2,3}\varepsilon_{ijk} G_k, \ \mbox{~where~~}    i,j=1,2,3,\\
\{G_{3+i}, \, G_{3+j}\} &=& \sum_{k=1,2,3}\varepsilon_{ijk} G_{3+k}, \ \mbox{~where~~}    i,j=1,2,3,\\
\{G_{i}, \, G_{3+j}\} &=& 0, \ \mbox{~where~~}    i,j=1,2,3.
\end{array}
\end{equation}

\section{Restrictions on the rank}\label{rank}

Let some $3d$ mechanical system have the Hamiltonian $H$ and 6 first integrals
$G_\alpha$ satisfying the commutation relations Eq. (\ref{o3o3}).

Consider two
$6\times 6$ matrices: the Jacobi matrix $J$ with elements
\begin{equation}\label{J}
J_{\alpha}^{\beta}:=\frac {\partial G_\alpha}
{\partial z_\beta},
 \ \mbox{~where~~}  \alpha,\,\beta=1,\,...\,,\,6,
\end{equation}
and the matrix
$P$ with elements
\begin{equation}\label{P}
P_{\alpha\beta}:= \{G_\alpha,\, G_{\beta}\},
 \ \mbox{~where~~}  \alpha,\,\beta=1,\,...\,,\,6.
\end{equation}
Then definitions  (\ref{J}) of Jacobi matrix and  (\ref{PB}) of brackets  imply that
\begin{equation}\label{PJ}
P_{\alpha\beta}=\sum_{\gamma,\delta=1, \dots, 6} J_{\alpha}^{\gamma} \omega_{\gamma\delta} J_{\beta}^{\delta}\,.
\end{equation}

Suppose that $G_1^2 +G_2^2 +G_3^2  \ne 0$ and $G_4^2 +G_5^2 +G_6^2  \ne 0$. Then the matrix $P$ has two independent null-vectors
\begin{equation}
(G_1,\,G_2,\,G_3,\,0,\,0,\,0)\ \ \mbox{and}\ \ (0,\,0,\,0,\,G_4,\,G_5,\,G_6)
\end{equation}
due to relations (\ref{o3o3}), and so $\mathrm{rank}(P)=4$.

Since the symplectic form $\omega$ is non-degenerate, the relation Eq. (\ref{PJ})
and degeneracy of the matrix $P$ imply that
\begin{equation}\label{ra}
\mathrm{rank}(P) \le \mathrm{rank}(J) <6.
\end{equation}

So either $\mathrm{rank}(J)=4$, or $\mathrm{rank}(J)=5$. Both these cases can be realized:
$\mathrm{rank}(J)=5$ for the Coulomb system
while $\mathrm{rank}(J)=4$ for some of the systems described in Section \ref{sec6}.

\section{Not all $3d$ systems have $\mathfrak{o}(4)$ symmetry}\label{omega}

To give an example of a $3d$ mechanical system without $\mathfrak{o}(4)$ symmetry%
, consider the Hamiltonian
\begin{equation}
H=H_1 + H_2 + H_3, \ \ \mathrm{where} \ \ H_i=\frac 1 2 p_i^2 + \frac {\omega_i^2} 2 q_i^2
\end{equation}
and where the $\omega_i$ for $i=1,2,3$ are
incommensurable.

Evidently, each of the functions $H_i$ is a first integral.

Let us show that each first integral of this system is
a function of the $H_i$, where $i=1,2,3$. Indeed, let $F$ be a first integral.
So, $F$ is constant on every trajectory defined for the system under consideration
by relations
\begin{equation}\label{tra}
q_i = \frac {\sqrt 2}{\omega_i} r_i \sin (\omega_i t +\varphi_i),     \ \ \ \ %
p_i = \sqrt 2 r_i \cos (\omega_i t +\varphi_i) \ \ \mbox{~~for~~}  i=1,2,3,
\end{equation}
where the $r_i$ and $\varphi_i$ are constants specifying the trajectory.
Since every trajectory given by Eq. (\ref{tra}) is everywhere dense on the torus
\begin{equation}
T(r_1,r_2,r_3):= \left\{z\in \mathbb R^6 \mid \frac 1 2 p_i^2 + \frac {\omega_i^2} 2 q_i^2 = r_i^2 \ \
\mbox{~~for~~}i=1,2,3\right\},
\end{equation}
it follows that $F$ is constant on every torus $T(r_1,r_2,r_3)$,
and  hence $F$ is a function of the $r_i$. This implies
$F=F(H_1, H_2, H_3)$.

Now suppose that the system has 6 first integrals $G_\alpha$ satisfying commutation
relations (\ref{o3o3}) of the Lie algebra  $\mathfrak{o}(4)$.
Then, since $G_\alpha=G_\alpha(H_1,H_2,H_3)$, it follows that the Jacobi matrix $J$ in Eq. (\ref{J})
is of rank $\leq 3$, and so due to Eq. (\ref{PJ}) the
matrix $P$, see Eq. (\ref{P}),
is of rank $\leq 3$. But this fact contradicts
the easy to verify fact that if $G_1^2+G_2^2+G_3^2\ne 0$ and $G_4^2+G_5^2+G_6^2\ne 0$,
then $\mathrm{rank}(P)=4$.

So, the system under consideration has no $\mathfrak{o}(4)$ symmetry in any domain invariant
with respect to Hamiltonian flow generated by $H$.

\section{An example of non-Coulomb $3d$ mechanical system with $\mathfrak{o}(4)$ Lie algebra of the first integrals}
\label{sec6}

Consider a particle in an homogeneous field  with potential $-q_{3}$.
This is a system with 3 degrees of freedom with Hamiltonian
\begin{equation}
H=\frac{\mathbf{p}^{2}}{2}-q_{3}  \label{H2}.
\end{equation}%

Let
\begin{equation}
U:=\{z\in \mathbb{R}^{6}\mid p_{1}^{2}<a_{1}^{2},\ p_{2}^{2}<a_{2}^{2}\},
\label{2}
\end{equation}%
where each $a_{s}$ is any smooth function of Hamiltonian $H$.
We denote the boundary of $U$ by $\partial U$ and its closure by $\bar U$.

Then the real functions
\begin{equation}
\begin{array}{l}
G_{1}=p_{1}, \\
G_{2}=\sqrt{a_{1}^{2}-p_{1}^{2}}\,\cos (q_{1}-p_{1}p_{3}), \\
G_{3}=\sqrt{a_{1}^{2}-p_{1}^{2}}\,\sin (q_{1}-p_{1}p_{3}), \\[2mm]
G_{4}=p_{2}, \\
G_{5}=\sqrt{a_{2}^{2}-p_{2}^{2}}\,\cos (q_{2}-p_{2}p_{3}), \\
G_{6}=\sqrt{a_{2}^{2}-p_{2}^{2}}\,\sin (q_{2}-p_{2}p_{3}),%
\end{array}
\label{FL}
\end{equation}%
are the first integrals defined in $\bar U$ and smooth in $U$.
Let $\mathcal{A}$ be
the space generated by $G_\alpha$.
The space $\mathcal A$, with Poisson brackets as an additional operation,
is the Lie algebra isomorphic to $\mathfrak{o}%
\left( 4\right) $.
It is subject to a direct verification that the integrals (\ref{FL}) indeed satisfy
the relations (\ref{o3o3})
for $\mathfrak o(4)$-generators.

The Casimirs, defined by the formulas
\[
K_1:=\sum_{i=1,2,3} G_i^2, \qquad K_2:=\sum_{i=1,2,3} G_{3+i}^2
\]
are equal to
\[
K_1=a_1^2, \qquad K_2=a_2^2
\]
and do not define the Hamiltonian only if
the $a_s$ are constant. In the case where the $a_s$ are constant, the Jacobi
matrix for the functions (\ref{FL}) has rank 4 at the generic point.
Otherwise, ${rank}(J)=5$ at the generic point.

\subsection{\textbf{Non-}Invariance of the domain $U$ under the flows $\mathcal L_{G}$%
.}

For $\lambda _{2}$ and $\lambda _{3}$ real, such that
$\protect\sqrt{\protect\lambda _{2}^{2}+\protect\lambda %
_{3}^{2}}=1,$ $\protect\lambda _{2}=\cos \protect\varphi ,$ \ $\protect%
\lambda _{3}=-\sin \protect\varphi$, we see that
$G:=\lambda G_{1}+\lambda
_{2}G_{2}+\lambda _{3}G_{3}$ is of the shape
\begin{equation}
\begin{array}{l}
G=\lambda p_{1}+Q\cos (q_{1}-p_{1}p_{3}+\varphi ),\\
\mbox{where } Q:=\sqrt{a_{1}^{2}-p_{1}^{2}}.
\end{array}
\end{equation}%

Set
\[
Q_H := \frac {dQ}{dH}=\frac {a_1}{Q} \frac {da_1}{dH}
\]
so that
\[
\{z_\alpha,\, Q \}=Q_H \{z_\alpha,\, H \} - \frac { p_1}{Q} \{z_\alpha,\,
p_1\},
\]
\[
\{z_\alpha,\, H \}=\sum_i \{z_\alpha,\, p_i \}p_i - \{z_\alpha,\, q_3 \}.
\]

Introduce a new variable $u$ instead of $q_{1}$:
\begin{equation}
u:=q_{1}-p_{1}p_{3}+\varphi .
\end{equation}%

Let $z(\tau_0)\in U$.
The equations of the Hamiltonian flow $\mathcal L_{G}$ are then of the form
\begin{eqnarray}
\frac{d}{d\tau }z_{\alpha } &=&\{z_{\alpha },G\},\ \ i.e.,  \nonumber \\
\frac{d}{d\tau }p_{3} &=&Q_{H}\cos (u)  \nonumber \\
\frac{d}{d\tau }q_{3} &=&Qp_{1}\sin (u)+ Q_{H}p_{3}\cos (u)  \nonumber
\\
\frac{d}{d\tau }p_{2} &=&0,\qquad \frac{d}{d\tau }q_{2}=Q_{H}p_{2}\cos (u)
\label{eq} \\
\frac{d}{d\tau }p_{1} &=&Q\sin (u),  \nonumber \\
\frac{d}{d\tau }q_{1} &=&\lambda +Qp_{3}\sin (u)-\frac{p_{1}}{Q}\cos
(u)+Q_{H}p_{1}\cos (u).  \nonumber
\end{eqnarray}

Since $\{G,H\}=0$, it is clear, that $dH/d\tau =0$ and $da_s/d\tau=0$
along the trajectories $z(\tau)$ defined by Eqs (\ref{eq}).


\newcounter{proposition}
\newcommand{\proposition}{\par\refstepcounter{proposition}%
{\bf Proposition \arabic{proposition}. }}

\newcounter{comment}
\newcommand{\comment}{\par\refstepcounter{comment}%
{\bf Comment \arabic{comment}. }}

\newenvironment{proof}[1][Proof]{\noindent\textsf{#1.\ }}{ \hfill \rule{0.5em}{0.5em}}


\proposition\label{prop}
{\it For any $z(\tau_0)\in U$, there exists a first integral $G_{z}\in \mathcal{A}$ such
that the Hamiltonian flow $\mathcal{L}_{G_{z}}$ leads the point $z(\tau_0)$ to the
boundary of $U$ for a finite time.}

\begin{proof}
We have
\begin{eqnarray}
\frac{d}{d\tau }u &=&\lambda -\frac{p_{1}}{Q}\cos (u), \\
\frac{d}{d\tau }p_{1} &=&{Q}\sin (u),            \label{p1dot}          \\
\frac{d}{d\tau }Q &=&-{p_{1}}\sin (u),
\end{eqnarray}%
and hence
\begin{eqnarray*}
\frac{d^{2}}{d\tau ^{2}}p_{1} &=&-p_{1}\sin ^{2}(u)+Q\cos (u)(\lambda -\frac{%
p_{1}}{Q}\cos (u))= \\
&=&-p_{1}+\lambda Q\cos (u)
\end{eqnarray*}

Further on we consider only the case $\lambda =0$. In this case
\begin{equation}
\frac{d^{2}}{d\tau ^{2}}p_{1}=-p_{1}.
\end{equation}%
and
\begin{equation}\label{p1}
p_{1}=p_{1}^{\mathrm{max}}\sin (\tau +\psi ),
\end{equation}%
where $p_{1}^{\mathrm{max}}\geq 0$ and $\psi $ are constant on the trajectories.

 We have
\begin{eqnarray}
\left( p_{1}^{\mathrm{max}}\right) ^{2} &=&p_{1}^{2}+\left( \frac{d}{d\tau }%
p_{1}\right) ^{2}=p_{1}^{2}+(a_{1}^{2}-p_{1}^{2})\sin ^{2}(u)  \nonumber \\
&=&a_{1}^{2}\sin ^{2}(u)+p_{1}^{2}\cos
^{2}(u)=a_{1}^{2}-(a_{1}^{2}-p_{1}^{2})\cos ^{2}(u)  \nonumber
\end{eqnarray}%
and
\be\label{32p}%
\begin{array}{l}
\left( p_{1}^{\mathrm{max}}\right) ^{2}=a_{1}^{2}-(a_{1}^{2}-p_{1}^{2}(\tau
))\cos ^{2}(u(\tau ))%
\end{array}%
\ee
for any $\tau$ since $p_{1}^{\mathrm{max}}$ is constant on each trajectory.

If $\cos(u(\tau_0))\ne 0$ and $p_1^2(\tau_0)<a_1^2(\tau_0)$, then
\be\label{33p}
\begin{array}{l}
\left( p_{1}^{\mathrm{max}}\right) ^{2}=a_{1}^{2}-(a_{1}^{2}-p_{1}^{2}(\tau
_{0}))\cos ^{2}(u(\tau _{0}))<a_{1}^{2}.%
\end{array}%
\ee
Eqs (\ref{33p}) and 
(\ref{p1}) imply that
\be\label{ineq}
p_1^2(\tau)<a_1^2(\tau) \ \ \mbox{for any $\tau$}
\ee
i.e., $z(\tau)\in U$ for any $\tau\in \mathbb R$.
Besides, conditions (\ref{32p}) and (\ref{33p}) imply
that
$$
cos(u(\tau)) \ne 0\ \ \mbox{for any $\tau$}.
$$

Now, observe that for every $z(\tau_0)$ it is possible to
choose $\lambda _{2}$ and $\lambda _{3}$
(i.e., $\varphi$) such that $\cos (u(\tau
_{0}))=0.$
Then, for this $\varphi$, we have $\left( p_{1}^{\mathrm{max}}\right) ^{2}=a_{1}^{2}$ and $Q(\pi/2
-\psi )=0,$
i.e., $z(\pi/2 -\psi )\in \partial U.$
\end{proof}

\comment
The proof of Proposition \ref{prop} shows also that for each fixed $\varphi$, the domain
\be
U_\varphi := \{z\in U \mid \cos(q_1-p_1 p_3 + \varphi)\ne 0 \}
\ee
is invariant under the action of Hamiltonian flow
$\mathcal L_{Q\cos (q_{1}-p_{1}p_{3}+\varphi )}$ acting
on $U_\varphi$ as 1-dimensio\-nal Lie group.

\comment
There is no domain $U_{common}\subset U$ invariant under Hamiltonian flows
$\mathcal L_{Q\cos (q_{1}-p_{1}p_{3}+\varphi )}$ for all $\varphi\in [0,\,2\pi)$.

Indeed,
$U_{common}\subset \bigcap_\varphi U_\varphi$,
and $\bigcap_\varphi U_\varphi = \varnothing$
since for any $z\in U$ there exists $\varphi\in [0,\,2\pi)$ such that
$ \cos(q_1-p_1 p_3 + \varphi) = 0 $.

\section*{Acknowledgements}
Authors are grateful to I.V.Tyutin and A.E.Shabad for useful discussions.
S.K. is grateful to Russian Fund for Basic Research
(grant No.~${\mathrm{17-02-00317}}$)
for partial support of this work.
S.B. was supported by the grant NYUAD-065.

\end{document}